\documentclass[preprint,showpacs,preprintnumbers,review,amsmath,amssymb,prb]{revtex4}

\usepackage{graphicx}
\usepackage{dcolumn}
\usepackage{bm}
\usepackage{epsfig}
\usepackage{wasysym}
\usepackage{amsmath}

\begin{document}
\preprint{PREPRINT}



\title{Isotropic soft-core potentials with two characteristic length
  scales and anomalous behaviour}


\author{Pol Vilaseca and Giancarlo Franzese}
\affiliation{Departament de F\'{\i}sica Fonamental, Facultat de F\'{\i}sica,
Universitat de Barcelona, Diagonal 647, 08028, Barcelona, Spain.}
\email{pvilaseca@correu.ffn.ub.es, gfranzese@ub.edu}

\begin{abstract}
Isotropic soft-core potentials with two characteristic length scales
have been used since 40 years to describe systems with polymorphism.
In the recent years intense research is showing that these potentials
also display polyamorphism and several anomalies, including
structural, diffusion and density anomaly. These anomalies occur in a hierarchy
that resembles the anomalies of water. However, the absence of
directional bonding in these isotropic potentials makes them different
from water. Other systems,
such as colloidal suspensions, protein solutions or liquid metals, can
be well described by these family of potentials, opening the
possibility of studying the mechanism generating the polyamorphism and
anomalies in these complex liquids.
\end{abstract}


\maketitle

\section{Introduction}
\label{}

Isotropic pair interaction potentials are usually considered the prototype for
simple atomic systems, such as argon. The most famous among them
is the phenomenological potential proposed in 1931 by Sir John Edward
Lennard-Jones (LJ) \cite{LJ}, commonly adopted as the textbook model
for real gases. The LJ potential incorporates the short--range
repulsion, due to the Pauli's quantum exclusion principle among
electron orbitals, as a function $\sim 1/r^{12}$ of the
distance $r$ between the centres of mass of the atoms. It also includes the
van der Waals attraction, due to instantaneous 
induced dipole-dipole London dispersion forces between the electron
clouds, as a long-range function $\sim -1/r^{6}$. These two
components are enough to generate a phase diagram 
with a gas, a liquid and a solid phase, as for neutral atoms or simple
molecules. The LJ model reproduces not only the thermodynamics, but
also the dynamics and the kinetics of these systems, providing for
example a good starting point for studying processes such as the
homogeneous nucleation of the crystal phase. 

The LJ model, and similar
potentials such as square wells, are useful also for more complex
systems, e. g.  colloidal suspensions
or protein solutions. In these cases these isotropic pair interaction potentials can
be used to represent  the interactions between the particles of the solute 
when the degrees of freedom of the solvent are
implicitly  taken into account in the effective interaction
potential.
However, there are (anomalous) properties of these and other systems, e. g. 
liquid metals or water, that
cannot be reproduced by simple potentials.  
It is, therefore, natural to ask if the family of isotropic potentials
can be extended in such a way to describe the phase diagram of the
anomalous substances.

Anomalous systems such as water and silica are network-forming liquids
with strongly anisotropic interactions. However, for other systems such as
liquid metals
\cite{SY76,YoungAlder77,KS77,LW77,KS78,MAC79,Cu81,VPRL83,KH84,DRB91,DebMetaLiq,SH93,Ve00}, colloids
\cite{BCESB00,Likos2001,
Quesada2001,Yethiraj03,BAKSH04} or
biological solutions \cite{protein,Vekilov2008,Buldyrev2008}
the use of soft-core
isotropic potentials with two characteristic length scales is a particularly suitable way of
constructing effective pair interactions capable of
describing the anomalies of these systems. Without the pretension to 
complete  or exhaustive coverage,
here we recall some recent results
 about this topic. For other aspects related to soft-core potentials we remit to 
previous reviews \cite{Bu02,Malescio07,KumarFBS2008}.

\section{Anomalous liquids}

Experiments for Ga \cite{MAC79}, Bi \cite{LSK76}, 
Te \cite{Th76}, S \cite{Sa67,Ke83}, Be, Mg, Ca, Sr, Ba \cite{Wa00},
SiO$_2$, P, Se, Ce, Cs,  Rb, Co, Ge \cite{DebMetaLiq},
Ge$_{15}$Te$_{85}$ \cite{Ts91}
and simulations for SiO$_2$ \cite{An00,Ru06b,Sh02,Po97}, S
\cite{Sa03} and BeF$_{2}$ \cite{An00} reveal the presence of a
temperature of maximum density (TMD) at constant pressure below which
the density decreases when the temperature is lowered
isobarically. This behavior is at variance with that of normal (or
argon-like) liquids where the density monotonically increases when the
temperature is decreased at constant pressure. The most famous
example of liquid with this anomaly is water, whose TMD at 1 atm is
at 4$^o$C. Below the TMD the isobaric thermal expansion coefficient
$\alpha_{P}\equiv\frac{1}{V}\left.\frac{\partial V}{\partial
    T}\right|_{P}$ of water assumes negative values. In normal liquids
$\alpha_{P}$ is always positive because it is proportional
to the (always positive) cross-fluctuation of volume and entropy.
Other thermodynamic anomalies of
water include the anomalous increase of isothermal volume
fluctuations, proportional to the isothermal compressibility 
$K_{T}\equiv\frac{1}{V}\left.\frac{\partial V}{\partial
    P}\right|_{T}$, below 46$^{o}$C at 1 atm and the anomalous
increase of isobaric entropy fluctuations, proportional to the
isobaric heat capacity $C_{P}\equiv T\left.\frac{\partial S}{\partial
    T}\right|_{P}$, below 35$^{o}$C at 1 atm  \cite{FS07}.

Another anomaly observed in water and other liquid 
is related to the diffusion coefficient $D$, defined as
\begin{equation}
 D\equiv\lim_{t\rightarrow\infty}\frac{\langle\Delta
   r(t)^{2}\rangle}{2dt}
\label{e1}
\end{equation}
where $t$ is the time, $d=3$ is the dimension of the system, 
\begin{equation}
 \langle\Delta
 r(t)^{2}\rangle\equiv\langle[r(t_{o}+t)-r(t_{o})]^{2}\rangle
\label{e2}
\end{equation}
is the mean square displacement of a single particle, $t_o$ is any time at equilibrium and the average is over the
initial $t_o$ and over the particles in the system. In a normal liquid $D$ decreases when
density $\rho$ or pressure $P$ are increased. Anomalous liquids, instead, are characterized by 
a region of the phase diagram where $D$ increases when  $P$ is increased at constant temperature $T$. In the case of water, for example, experiments show that the normal behaviour of $D$ is restored only at
pressures higher than $P\approx1.1$ kbar at 283~K \cite{An76}. 

Also the structure can be anomalous. Normal liquids tend to become more
structured when compressed. This can be quantified by a
translational order parameter $t$ that measures the tendency of the
molecules to adopt preferential separations, and by an orientational 
order parameter $Q_{l}$ 
that measures the tendency of a molecule and its 
nearest neighbours to assume a specific
local arrangement, as considered by Steinhardt et al. \cite{St83}.
The translational order parameter is defined as \cite{Sh02,Er01,Er03}
\begin{equation}
 t\equiv\int_{0}^{\infty}|g(\xi)-1|d\xi
\label{e3}
\end{equation}
where $\xi\equiv r\rho^{1/3}$ is a reduced distance (in units of the
mean interparticle separation $\rho^{-1/3}$) and $g(\xi)$ is the
radial distribution function. As the parameter $t$ depends only on the deviations of $g(\xi)$ from
unity, its value is sensible to long range periodicities. For an ideal
gas $g(\xi)$ is constant and equal to 1 and there is no translational
order ($t=0$). For a crystal phase $g(\xi)\neq1$ for long distances
and $t$ becomes large.

The orientational order parameter is by definition \cite{St83}
\begin{equation}
 Q_{l}\equiv\frac{1}{N}\sum_{i=1}^{N}Q_{l}^{i} 
\label{e4}
\end{equation}
where $l=1,2,...$ is an index, $Q_{l}^{i}\equiv\left[\frac{4\pi}{2l+1}\sum_{m=-l}^{m=l}|(\bar{Y}^{i}_{lm})_{k}|^{2}\right]^{1/2}$, $k$ is a fixed number of nearest neighbour particles and
$(\bar{Y}^{i}_{lm})_{k}\equiv\frac{1}{k}\sum_{j=1}^{k}Y_{lm}(r_{ij})$
is the average of the spherical harmonics $Y_{lm}$ with indices $l$
and $m$, evaluated over the
vector distance $r_{ij}$ between particles $i$ and $j$. 
For $l=6$ and $k=12$, 
$Q_{6}$ reaches its minimum value
$Q_{6}^{\rm ih}=1/\sqrt{k}=0.287$ for an isotropic homogeneous system, 
while for a fully ordered
f.c.c arrangement is $Q_{6}^{\rm fcc}=0.574$.

For a normal liquid $t$ and $Q_{l}$ increase with
pressure. Anomalous liquids,  instead, show a region where the structural order parameters
 decrease for increasing pressure (or density) at constant $T$, i.e. the
 system becomes more disordered. This is what has been observed, for example, in
molecular dynamics simulations for water by  
Errington and Debendetti \cite{Er01} and by Shell et al. for silica \cite{Sh02}. 

All the anomalies of water have a well determined sequence found in
experiments \cite{An76} and simulations \cite{Er01}.
In the $T-\rho$ plane the water 
structural anomaly region is encompassing the diffusion anomaly region
which includes the density anomaly region. However in other
liquids the sequence of anomalies may be 
different. For example, for silica the anomalous diffusion region
contains the structural anomaly regions that, in turn, includes
the density anomaly region \cite{Sh02}.

\subsection{Polymorphism, Polyamorphism and Liquid--liquid phase transition}

Another anomaly that has
received considerable attention in recent years is 
the possible existence of a liquid--liquid (LL) phase transition for
single-component systems with a standard gas-liquid critical point. 
The two coexisting liquids, the high density liquid (HDL) and
the low density liquid (LDL), would differ in density and local
structure, as proposed by Poole et al. in 1992 \cite{Po92}, based on
simulations for a water model. At that time it was  known that
water can have more than an amorphous state \cite{Mishima1985}, a 
high density amorphous (HDA) and a low density amorphous (LDA),
separated by discontinuous density--change with the character of
first-order phase transitions. An even higher amorphous state, the very
HDA (VHDA), of water has been observed in 2001 by Loerting and
coworkers \cite{VHDA}.

This property, called polyamorphism, i.e. the occurrence of more than one
amorphous state, is usually associated to the polymorphism, 
i.e. the occurrence of more than one crystal phase \cite{polymorphysmRev}.
Typical examples of polymorphic substances are
water with at least 16 forms of ice, and carbon with 
diamond and graphite (made of graphene sheets, as are fullerenes and carbon
nanotubes).

Direct evidences for liquid polyamorphism
have been observed experimentally in phosphorous
\cite{Ka00,Ka04,Mo03} in 2000, triphenyl phosphite
\cite{ReiKurita10292004,Ta04,Kur05} in 2004 and in yttrium oxide-aluminum
oxide melts \cite{Greaves08} in 2008.
Experimental data consistent with a LL phase transition
have also been presented for single-component systems, besides water,
such as silica \cite{An96,Po97,La00}, carbon \cite{Th93}, 
selenium \cite{Br98}, and cobalt
\cite{Va03}, among others \cite{Aa94,Wi01,WBM02}.  

A LL critical
point has been predicted by simulations for
all the commonly used models of water, including SPC/E, ST2, TIP4P
and TIP5P \cite{Po92,brovchenko05}, and for
 specific models of
phosphorous \cite{Mo01}, supercooled silica \cite{An96,Po97,Vo01b,Sa03},
hydrogen \cite{Sca03}, and carbon based on molecular dynamics
simulations \cite{Gl99}, For carbon, however,
subsequent ab initio simulations \cite{Christine02} and simulations
for a semiempirical potential partly based on {\it ab initio} data
\cite{Gh05} did not confirm this finding.

\section{Isotropic models with anomalies}

The anomalies described in the previous section can be reproduced by
using isotropic core-softened potentials.
 The  core-softened potentials are usually characterized by a change
 of curvature within the repulsive range, such as a ramp or
a shoulder.

\begin{itemize}
\item Ramp-like potentials have their softened region defined by a
  repulsive ramp,
   establishing two competing equilibrium distances, and in some cases
  an attractive well \cite{HemmerStellA,St72,Ja98,Ja99a,Ja99b,Ja01a,Ja01b,Wi02,Ku05,Yan05,Yan06,Yan08,Wi06,Ol06a,Ol06b}.

 \item Shouldered potentials are composed by a
   hard-core, the repulsive shoulder softening the core, and an
   optional attractive well
   \cite{Ki76a,DRB91,Ki76b,Ha73,Ve00,La98,La99,Sc00,Sc01,Fra01,Ma05,Bu02,Bu03,BB04,Ol05,Camp03,Camp05,Wi02,Fr07a,OFNB08,GFFR09}.

\end{itemize}

Thermodynamic anomalies have been widely reported
  for such potentials. Some of these anomalies depend on the details of
  the potentials, as we will show in the following.
 Among 
   these potentials, those with an attractive part display more than
   one first-order phase transition.

It was after almost 40 years from the introduction of the LJ potential
that Hemmer and Stell explored the idea of
core-softening, often referred as core-collapse, proposing the first
repulsive-ramp potential
\cite{HemmerStellA,St72}. 
They realised that by 
making the core more penetrable they could induce a second 
first-order phase transition in systems that already have one.  
They justified the occurrence of the second transition in a lattice
gas by means of a heuristic argument based on the (particle-hole)
symmetry between occupied and unoccupied cells.
This argument, however, is not applicable to continuous systems where
the particle-hole symmetry does not hold. Nevertheless, they showed
explicitly for fluids in one dimension how a similar phenomenon could
appear by softening the hard-core, finding a range of parameters for
which the phase diagram displays two first order phase transitions,
both ending in critical points. 
The kind of potentials they proposed were discontinuous. However, they
argued that other analytical potentials arbitrarily close to those
they considered could be constructed and the second transition would
persist for such potentials. 
The high-density critical point was first interpreted as a solid-solid
isostructural phase transition for systems such as Cs or Ce \cite{BeeSw}.
In their original work \cite{HemmerStellA,St72} Hemmer and Stell 
remarked that for one dimensional
models with long range attraction 
the isobaric thermal expansion coefficient
$\alpha_{P}$ 
can take anomalous
negative values. 

In 1976, Kincaid et al. \cite{Ki76a,Ki76b} proposed a variant of the
Hemmer--Stell potential with  
the hard core softened by adding a
finite shoulder of constant positive energy. They  found at high density
and pressure an isostructural phase transition. Kincaid and Stell
\cite{KS77} extended this work by 
introducing a shouldered potential without attraction
to describe solid mixtures with isostructural transitions.

 After this pioneering works, several soft-core potentials have been
 proposed and analyzed with approximate methods or numerical
 simulations to study the properties of complex liquids as 
 liquid metals, alloys, electrolytes, colloids and, to some extent, water
 \cite{SY76,YoungAlder77,KS77,LW77,KS78,MAC79,Cu81,VPRL83,KH84,BCESB00,Likos2001,DRB91,DebMetaLiq,SH93,HG93,SS97,HydraProBi,RobSinZhuEv}. 

For example, Stillinger and Weber \cite{SW78} in 1978 used
molecular dynamics simulations to study the phase behaviour of the
Gaussian core model. They reported the surprising and unexpected
finding that the model at equilibrium displays water-like anomalies,
such as a density of maximum melting temperature, hence a region of
decreasing volume upon melting, a negative thermal expansion coefficient
in the fluid phase and an increase of self-diffusion upon isothermal
compression.  

Later, in 1991, Debenedetti et al. \cite{DRB91} showed that anomalies
can occur in two-length-scales potentials also when the inner distance
is attractive and the larger distance is repulsive. They showed that
such a potential on a lattice 
can induce the formation of an open (low density)
structure at low $P$ and $T$. Upon heating or pressurization
the open structure looses stability and collapses into
a closed (denser) structure, leading to the anomaly in density and to
negative $\alpha_{P}$. 
A similar model on a lattice was studied by de
Oliveira and Barbosa \cite{Ol05} finding a LL coexistence with a line
of critical points.
More recently Archer and Wilding \cite{Archer07} studied a different
isotropic model with short range attraction and larger range
repulsion, supporting the existence of  a similar line of critical points.

In 1993 Head-Gordon and Stillinger \cite{HG93,SH93} showed that from
the inversion of the radial distribution function of water is possible
to deduce an effective oxygen-oxygen interaction potential that
resembles a continuous version of  the Stell--Hemmer shouldered potential.
In 1996, Cho et al. \cite{Ch96b,Ch96a} extended even further
the original idea of Stell and Hemmer of a potential with two length scales, by
proposing a potential with an inner well and an outer well with two
characteristic attractive energies separated by a local maximum. They
analyzed the model in one 
dimension showing the presence of the density anomaly and proposed to
consider it as an effective potential for the second shell of water,
assuming that the first shell can be considered as part of an
invariant inner core that does not play a relevant role in the
anomalous density behavior of water.

In 1998, Sadr-Lahijany et al. \cite{La98,Sc01} 
considered a similar shouldered
potential in two dimensions, both in the discrete and the continuous
version, with a deep attractive well. The  
shoulder gives rise to an inner distance that is less attractive than
the outer distance.
They found polymorphism: a low-$P$ triangular lattice 
 less dense than the liquid,  and  a  high-$P$ square
lattice  denser than the liquid. 
Furthermore, they reported the occurrence of
three anomalies: the density anomaly, the increase of
isothermal compressibility upon cooling, and the diffusion anomaly.
By a simplified argument
in one dimension, it was shown how the increase of $P$ reduces the
Gibbs free energy  at the inner distance of the potential, with respect
to the larger and more attractive distance, inducing the collapse to
the denser structure.
The interplay between the two different local structures 
near the freezing line was proposed as the mechanism responsible of the anomalies. 
This rationalization was better clarified by Wilding and Magee \cite{Wi02}.
Further studies on a continuous version of the model by Netz et al. \cite{Ne04}
showed anomalous behavior in the stable region of the phase diagram if
the outer minimum is deeper than the inner minimum. In the case of a
deeper inner minimum, anomalous behavior occurs inside
the unstable region. Nevertheless, the study of the same model in
three dimensions at low $T$ in a stable liquid state by Quigley and
Probert \cite{Qui05} suggests that the anomalous behavior of such model
is unique to the two-dimensional case.

Also in the 1998, Jagla \cite{Ja98,Ja99a} studied a version of the 
purely repulsive Kincaid and Stell potential by softening the core 
with a linear ramp, as in the original Hemmer and Stell potential.
The absence of the attractive term implies the absence of
the liquid-gas phase transition. Under pressurization the potential displays many 
crystalline polymorphs and anomaly in density. Also in this case
the occurrence of these anomalies is associated to the competition
between the two scales of the potential: the hard-core distance and
the distance at which the interaction becomes zero. 
The appearance of
complex ground state structures in the system is a consequence of the
competition between two terms in the enthalpy $H$: the pressure--volume
term $PV$ leads to minimize the volume, while the soft-core repulsive
term leads to maximize the
interparticle distance. The different arrangements of particles in the
ground state depend on the values of the pressure $P$ and the values of
the two length scales of the potential. This result was later
confirmed by Kumar et al. \cite{Ku05} using integral-theories. 
Polymorphism for purely repulsive core-softened potentials 
was also obtained a few years later by Velasco et
al. in collaboration with Hemmer and Stell  \cite{Ve00,He01}.
More recently, by using free-energy calculations, Gribova et al. \cite{GFFR09} 
showed that for this potential the system
exhibits the water-like density and
diffusion anomaly and that  the anomalies move to the
region where the crystal is stable with increasing 
repulsive-step width.

In 1999, Jagla  analyzed  a slightly modified version of  
the Hemmer and Stell potential
\cite{Ja99b} in three dimensions showing by
simulations the
appearance of water-like anomalies and predicting by theory a LL
critical point. However, in the simulations the LL critical point
results inaccessible due to the inevitable crystallization
\cite{Ja99b}. 

In 2001,
by adopting different parameters for the 
potential \cite{Ja01a} and by analyzing a continuous version of the
model \cite{Ja01b},
Jagla showed that the LL phase transition is observed in simulations in
two and three dimensions, when the attraction is strong enough, while
disappears if the attraction strength is approaching zero
\cite{Ja01b}. In both case, however, the density displays hysteresis
at low $T$ due to mechanical metastabilities, and water-like anomalies
are observed, regardless the absence of the LL critical point, showing
that the presence of the anomalies is not a sufficient condition for
the occurrence of the LL critical point \cite{Ja01b}.
Later, Wilding and Magee observed that
the locus of density maxima ends very close to the LL
critical point, when this is present \cite{Wi02}. 
Xu et al. \cite{Xu05,Xu06} showed
that in the super-critical region of the LL critical point 
the dynamics changes from Arrhenius to non-Arrhenius when the line of
maxima correlation length, approximated with the line of maxima of the
specific heat $C_P$, is crossed.
Caballero and Puertas \cite{Ca06} studied the model by first-order
perturbation theory for different choices of the attractive 
range, finding that the LL phase transition is 
attraction-driven for long ranged potentials, and is
compression-driven when the interaction is shortened.  

In the same year 2001, Franzese et al. \cite{Fra01} showed
that the occurrence of the anomalies is not a necessary condition for
the LL critical point. They found, by simulations
and integral equation theory \cite{Fra02},  that a
discontinuous shouldered well (DSW) potential in three dimensions,
similar to that originally proposed  by Kincaid et al.,
displays a LL coexistence, ending in a critical point, metastable with
respect to the crystal. They also showed that the system does not display
an anomalous behavior in density. 
This result
suggested that the type of systems displaying the LL phase
transition could be broader than what was previously hypothesized, and
 that experimental evidences of a LL phase transition
should be seeked also in systems without density anomaly. 
In 2004 Skybinsky et al. \cite{Sk04}
showed that, by changing the parameters of the DSW, the phase diagram
displays a LL critical point that is stable with respect the
spontaneous homogeneous crystallization. They  developed a modified
van der Waals equation that
qualitatively reproduces the behavior of both liquid-gas and LL
critical points of the model. In 2005 Malescio et
al. extended the previous work 
to the case
of large soft-core ranges, by using an integral equation approach in
the hypernetted-chain approximation. They 
showed that only a limited range of parameters of the DSW model give
rise to a phase diagram with an accessible LL critical point
and that this occurs when the repulsive component of the potential
equilibrates the attractive component, in particular when the
repulsive volume weighted by the repulsive energy compensates the
attractive volume weighted by the attractive energy \cite{Ma05}. 
Lattice Monte Carlo simulations by Balladares and Barbosa \cite{BB04}
for the DSW confirmed the LL coexistence but gave a phase
diagram with a line of critical points connecting the  LL phase
transition and the liquid-gas phase transition. This
feature appears to be an artifact of the lattice, because
off-lattice Monte Carlo simulations by Rzysko et al. \cite{rzysko08}
showed no such a line of critical 
points, confirming the previous results of molecular dynamics
simulations and theory \cite{Fra01,Fra02,Sk04,Ma05} and providing an
accurate estimates of the LL critical points and its exponents.

In 2003, Buldyrev and Stanley \cite{Bu03} tested the idea that for any
characteristic length scale in the interaction potential the system
possibly 
displays a different liquid phase and a new LL phase transition.
They added an extra discontinuous step to the soft-core of 
the DSW, defining a third characteristic distance. 
For certain values of the parameters of the potential the system
presents up to three first-order phase transitions between fluids of
different densities ending in critical points. 
The radial
distribution functions $g(r)$ 
displays dramatic differences in structure between the low density
liquid (LDL) and
the high density liquid (HDL), but less pronounced differences between
the HDL
and the very high density liquid (VHDL). 
These results suggest 
that more critical points could be created by adding more steps to the
potential and carefully selecting the parameters. However, for
$k_{B}T$ larger than the steps of the potential, the effect of the
new length scales becomes
negligible and the phase diagram converges to that  of a system with
a continuous potential.

Nevertheless, a few years later Cervantes et al. \cite{cervantes07}
showed, by calculating the free energy by
discrete perturbation theory, that there is a
range of combination of parameters of the two-scales DSW that generate a phase
diagram with three critical points: between gas and liquid, between
LDL and HDL, and between HDL and VHDL.
The three critical points for the two-scales DSW were found  also,  by Artemenko et
al. \cite{Artemenko08}, by
analyzing the modified van der Waals equation proposed in Ref.~\cite{Sk04}.
Therefore is not necessary to add more than two 
length scales in the potential to reproduce phase diagrams with more
than two fluid-fluid critical points. It is enough to have two
characteristic length scales that compete creating a multiplicity of
minima for the free energy at different $P$ and $T$.
These results, about the minimum conditions to reproduce 
polyamorphism including at least four different characteristic
densities, are particularly interesting considering that
experiments \cite{VHDA} and molecular models simulations \cite{brovchenko05}
show the existence of a very high density amorphous phase of water
\cite{Loerting_Giovambattista}. 

Polyamorphism and polymorphism were also studied in two dimensions in the
Kincaid-Stell purely repulsive shouldered model by Malescio and
Pellicane  in 2003 \cite{MalPel2003} and in a purely repulsive
potential with a shoulder and a long-range repulsive tail by Camp
\cite{Camp03,Camp05}. The fluid phase at low temperature
exhibits a very rich variety of structures, including chains, stripes
and polygons \cite{duncan04}. The Camp model approximates a two-dimensional
system of dipolar particles in a strong field aligned perpendicular to
the plane \cite{Osterman07,Dobnikar08}.

The properties of the core-softened potentials depend on the ratio
between the two characteristic scales. Yan and co-workers
\cite{Yan05,Yan06} 
explored a range of systems going from a hard-sphere potential to a
pure ramp without hard-core, finding that thermodynamic (negative
$\alpha_P$) and dynamic (diffusion) anomalies occur almost across the
entire range, while water-like structural anomalies occur only
for cases with the ratio between the two length scales 
comparable with that between the first two peaks of water in standard conditions.
They also showed that the anomalies 
have the same hierarchy as in water:
the density anomaly occurs within the region of diffusion anomaly,
that is found within the region of structural anomaly, as observed for the
SPC/E water \cite{Er01}. This similarity with water exists despite the
lack of directionality in the isotropic potentials. 
The analogy with water was pushed forward  by Yan et al. \cite{Yan08}
comparing the anomalies in the super-critical region of the two scale
potential with those of TIP5P water, giving new support to the initial
proposal by Cho et al. \cite{Ch96a} that two-scale isotropic
potentials could represent an effective interaction for the
second shell of water, being the first part of the core.

The  water-like hierarchy was found also 
by de Oliveira et al. \cite{Ol06a,Ol06b} in three
dimensions for a shouldered potential with 
a small attractive region, that resembles the Camp potential
\cite{Camp03} and that has been extensively studied by Pizio et
al. recently \cite{Pizio2009a}.
These results on
all these soft-core models show that orientational interactions, such
as hydrogen bonding, are not a necessary condition for the presence of
water-like
anomalies. The anomalous features at about the temperature of the
maximum density are caused by the reduction of the large empty spaces
around the molecules upon compression and heating. This effect is
captured by the soft-core potentials with two characteristic length
scales. However, other anomalies of water at lower $T$, such as the
liquid-liquid phase transition and the associated Widom line are not
reproduced in the correct way. In particular their slope in the $P-T$
phase diagram is opposite to what is observed in molecular water
models such as TIP4P. This difference implies an opposite behaviour as
function of $P$ for thermodynamic and dynamic quantities in the
supercooled regime.

All the example of potentials mentioned above have been proposed for
systems going from water to colloids, to protein solutions, to liquid
metals, with specific experimental cases, such as 
polystyrene monolayers between
water and air in two dimensions  \cite{Quesada2001}, or
gallium
\cite{MAC79}, or tellurium \cite{Hafner90} or liquid alkaline-earth
metals near the melting point \cite{Wa00} in three dimensions. 
The effective (pseudo)potentials
representing these systems are often similar in shape, but with
different details. It is therefore interesting to understand how their
properties depend on the detailed features of the potential. 
To this goal we present in the next section a case study for a highly
tunable potential that offers the
opportunity to explore the properties of a large family of soft-core
potentials.  This
potential is a continuous version of the DSW, the continuous shouldered well (CSW) potential,
introduced by Franzese in 2007 \cite{Fr07a}. It was shown that the CSW
potential has density anomaly in three dimensions \cite{Fr07a} and a
detailed analysis by de Oliveira et al. \cite{OFNB08} revealed also
diffusion anomaly and structural anomaly with a water-like hierarchy.
Recently  Standaert et al. \cite{Standaert10} have adopted this
model to study the condition of anomalous (non-Gaussian) self-diffusion in a system
driven out of equilibrium by intermittent length rescaling.

\section{The Continuous Shouldered Well potential}

\begin{figure}
\begin{center}\includegraphics[clip=true,scale=0.375]{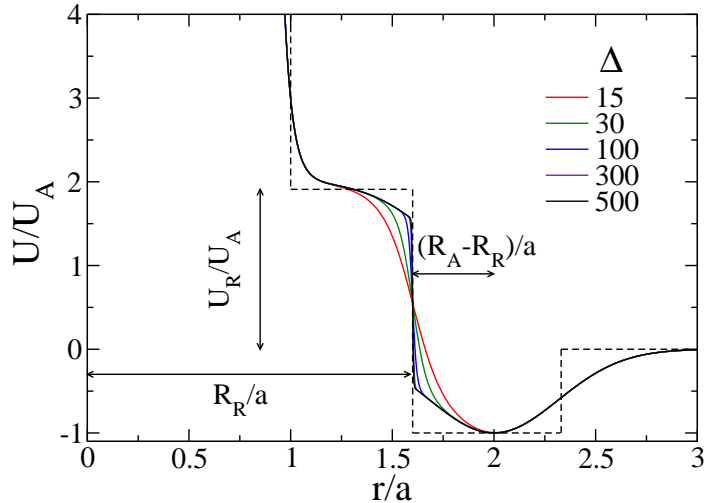}
\end{center}
\caption{The Continuous Shouldered Well (CSW) potential for 
$\Delta=15,30,100,300,500$ (continuous lines) and the
 Discontinuous Shouldered Well (DSW) potential (dotted black
  line).
By increasing $\Delta$ the
  CSW potential approximates the DSW around $R_{R}$.} 
\label{f1}
\end{figure}

The CSW model \cite{Fr07a} consists of a set of
identical particles interacting through the isotropic pairwise
potential 
\begin{equation}
 U(r)=\frac{U_{R}}{1+\exp\left(\Delta(r-R_{R})/a\right)}-U_{A}
 \exp\left[-\frac{(r-R_{A})^{2}}{2\delta_{A}^{2}}\right]+\left(\frac{a}{r}\right)^{24}\,
\label{e5}
\end{equation}
(Fig.~\ref{f1}) where $a$ is the diameter of the particles, $R_{A}$ and $R_{R}$ are
the distance of the attractive minimum and the repulsive
radius, respectively, $U_{A}$ and $U_{R}$ are 
the energies of the
attractive well and the repulsive shoulder, respectively, $\delta_{A}^{2}$ is the
variance of the Gaussian centered in $R_{A}$, and $\Delta$ is the
parameter which controls the slope between the shoulder and the well
at $R_{R}$.
Varying the parameters the potential can
be tuned from a repulsive shoulder to a deep double well. 
In particular, by increasing $\Delta$ the soft-core repulsion becomes
more penetrable near the minimum of the attractive well, and the
softness of the potential increases for $r>R_R$ and decreases for $r<R_R$.
We fixed here the set of values $U_{R}/U_{A}=2$, $R_{R}/a=1.6$, $R_{A}/a=2$, $\left(\delta_{A}/a\right)^{2}=0.1$, while we change $\Delta$, considering the values $\Delta=15,30,100,300,500$ (Fig.~\ref{f1}) going from the case
$\Delta=15$ studied in Ref.~\cite{Fr07a,OFNB08} to slopes
that approach the infinite value of DSW.

\section{The Phase Diagram}

\begin{figure}
\begin{center}\includegraphics[clip=true,scale=0.55]{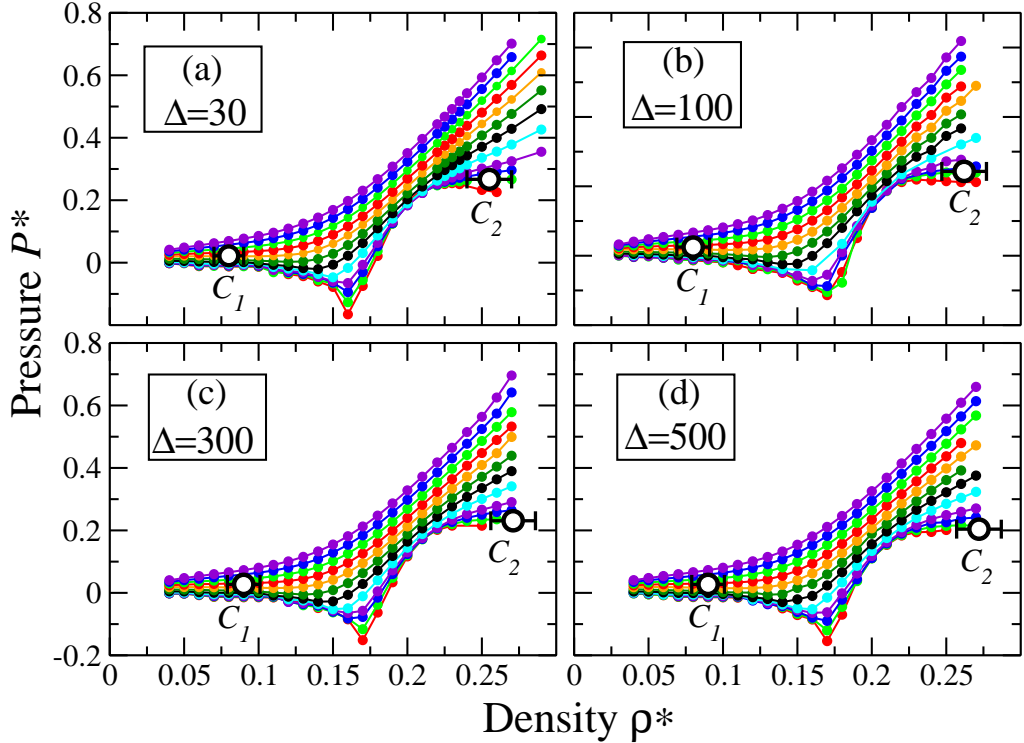}
\end{center}
\caption{Isotherms in the $P^{*}-\rho^{*}$ phase diagram from
  simulations for CSW potential for $\Delta=30$ (a), $\Delta=100$ (b), $\Delta=300$ (c) and $\Delta=500$ (d) (in all the panels, from top to  bottom, $T^{*}=1.4$, $1.3$, $1.2$, $1.1$, $1.0$,
  $0.9$, $0.8$, $0.7$, $0.6$, $0.55$, $0.5$ and $0.45$).} 
\label{f2}     
\end{figure}

\begin{figure}
\begin{center}\includegraphics[clip=true,scale=0.55]{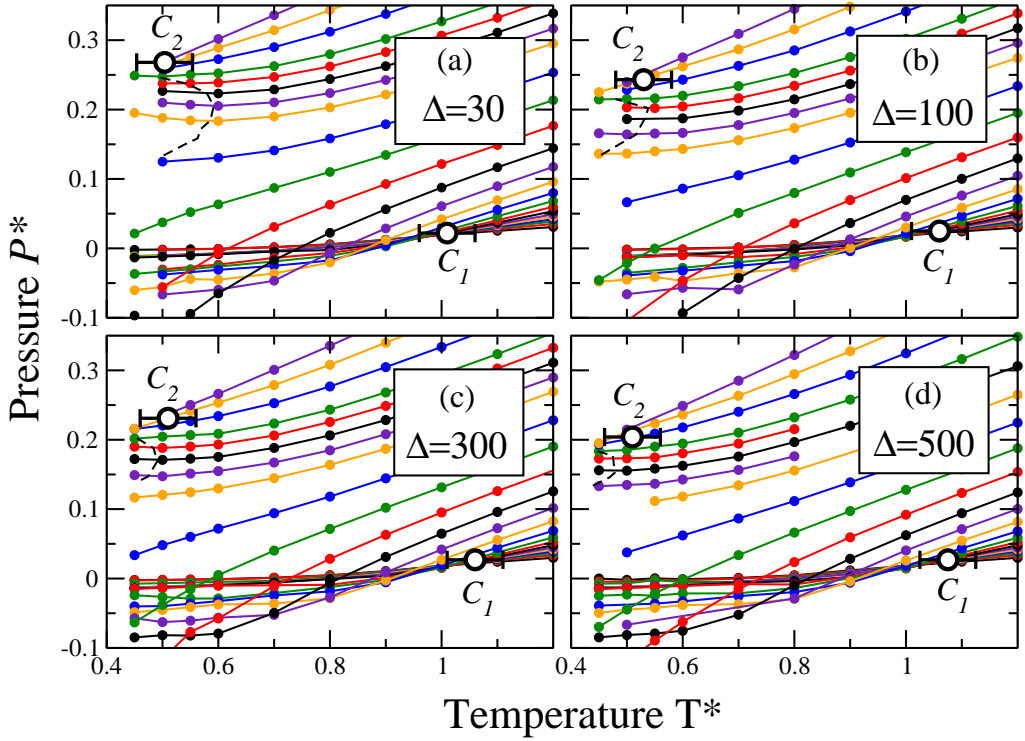}\end{center}
                \caption{The $P^{*}-T^{*}$ phase diagram for                  
$\Delta=30$ (a), $\Delta=100$ (b), $\Delta=300$ (c) and $\Delta=500$                  
(d). In all the panels, we show   (from bottom at $T^*=1.2$) 
isochores from $\rho^{*}=0.04$ to 
$0.20$ (with constant density separation $\delta \rho^*=0.01$), from $0.205$
to $0.215$ (with $\delta \rho^*=0.005$), and 
from $0.22$ to $0.25$ (with $\delta \rho^*=0.01$). The points
                  where isochores cross correspond to the gas-liquid and
                  LDL-HDL critical points ($C_1$ and $C_2$,                  
respectivelly). The black dashed line at low $T$ is a guide for
                  the eye estimating the TMD line, correspondig to the
                  line of minima along the isochores.} 
\label{f3}             
\end{figure}

The phase diagram for all the considered values of $\Delta$ display
the same qualitative behaviour in the plane
$P^{*}-\rho^{*}$ (Fig.~\ref{f2}) 
 where $P^{*}\equiv Pa^{3}/\epsilon$ and
$\rho^{*}\equiv\rho a^{3}$ are reduced pressure and density,
or $P^{*}-T^{*}$ (Fig.~\ref{f3}) where 
$T^{*}\equiv k_{B}T/\epsilon$, with $k_{B}$ Boltzmann
constant, is the reduced temperature. 
In particular, at low $T^{*}$ the isotherms are
non-monotonic (van der Waals loops), corresponding to the coexistence
of (i) gas and liquid at low $\rho^{*}$, (ii) 
low density liquid (LDL) and high density liquid (HDL) at higher
$\rho^{*}$. 
The two coexistence regions end in critical points:  $C_{1}$ for
gas-liquid coexistence,  $C_{2}$ for LDL-HDL coexistence. 
For $\Delta<100$ the HDL phase is metastable with respect to the crystal,
but with a lifetime long enough to allow us to equilibrate the liquid around $C_{2}$.
For $\Delta\geq100$ the HDL lifetime is too short to equilibrate the liquid around 
$C_{2}$ and we extrapolate the location of $C_{2}$ by linear fitting 
from higher $T^{*}$ isotherms. 

As  $\Delta$ is increased, both $C_{1}$ and $C_{2}$ tend to the
corresponding values for the DSW potential. This result is consistent
with the idea that the DSW can be seen as a limiting case of the
family of CSW potentials presented here. However, temperature and
pressure of the LDL-HDL critical point for $\Delta=500$ 
($T^{*}_{C_2}=0.52\pm0.01$, $P^{*}_{C_2}=0.204\pm0.007$,  
$\rho^{*}_{C_2}=0.272\pm0.008$)
are still far from
the values for the DSW potential 
($T^{*}_{C_2}=0.69\pm0.02$,
$P^{*}_{C_2}=0.110\pm0.002$, $\rho^{*}_{C_2}=0.280\pm0.020$).

\section{Anomalies}

The CSW has anomalies in density, diffusion and structure. By
numerical simulations, de Oliveira et
al. \cite{OFNB08} studied the case $\Delta=15$ finding the same
hierarchy of anomalies as reported for other two-scale potentials and
for the SPC/E water.  

In Ref.~\cite{Vilaseca10} we analyze in details the cases $\Delta=30,
100, 300, 500$. In summary, for all the considered values of $\Delta$, 
we observe the following.

\begin{itemize}
\item Density anomaly.

We find a temperature of
minimum pressure $\left(\partial P/\partial T\right)_{\rho}=0$ along
the isochores near the LDL-HDL critical point 
$C_{2}$. 
These temperatures correspond to the TMD line
at constant $P$, because the condition $\left(\partial P/\partial
  T\right)_{V}=0$ implies, according to Maxwell relations,
$\left(\partial S/\partial V\right)_{T}=0$. Consequently, it is
$\left(\partial S/\partial P\right)_{T}=\left(\partial S/\partial
  V\right)_{T}\left(\partial V/\partial P\right)_{T}=0$ and, again
using Maxwell relations, $\left(\partial V/\partial
  T\right)_{P}=\left(\partial \rho/\partial T\right)_{P}=0$, which
implies the presence of the temperature of maximum density (TMD) at
constant $P$.

\newpage
\item Diffusion anomaly.

We find  an anomalous-diffusion region, i.e. a
region 
($\rho_{Dmin}<\rho<\rho_{Dmax}$) 
where $D$, defined by Eq.(\ref{e1}),
increases with
increasing density at constant $T$.

\item Structural anomaly.

We find that $t$, defined as in Eq.(\ref{e3}) increases with increasing
density, for $\rho<\rho_{t_{\rm max}}$,
and reaches a maximum at $\rho_{t_{\rm max}}$. Above $\rho_{t_{\rm
    max}}$, for
increasing $\rho$,  $t$ decreases until it reaches a minimum at
$\rho_{t_{\rm min}}$. For $\rho>\rho_{t_{\rm min}}$, $t$ recovers the normal
behaviour.

Moreover, we observe that $Q_{6}$, as defined in Eq.(\ref{e4}),  has a
non--monotonic behaviour along the isotherms with a 
maximum at $\rho_{Q_{\rm max}}$.
The density $\rho_{Q_{\rm max}}$ for each isotherm lies between
$\rho_{t_{\rm max}}$
and $\rho_{t_{\rm min}}$. In the area between $\rho_{Q_{\rm max}}$ and
$\rho_{t_{\rm min}}$
both order parameters decrease for increasing $\rho$, hence the liquid
becomes more disordered for increasing density. This behaviour defines
the structural anomaly region ($\rho_{Q_{\rm
    max}}\leq\rho\leq\rho_{t_{\rm min}}$).

\item{Hierarchy of anomalies.}

\begin{figure}
\begin{center}\includegraphics[clip=true,scale=0.55]{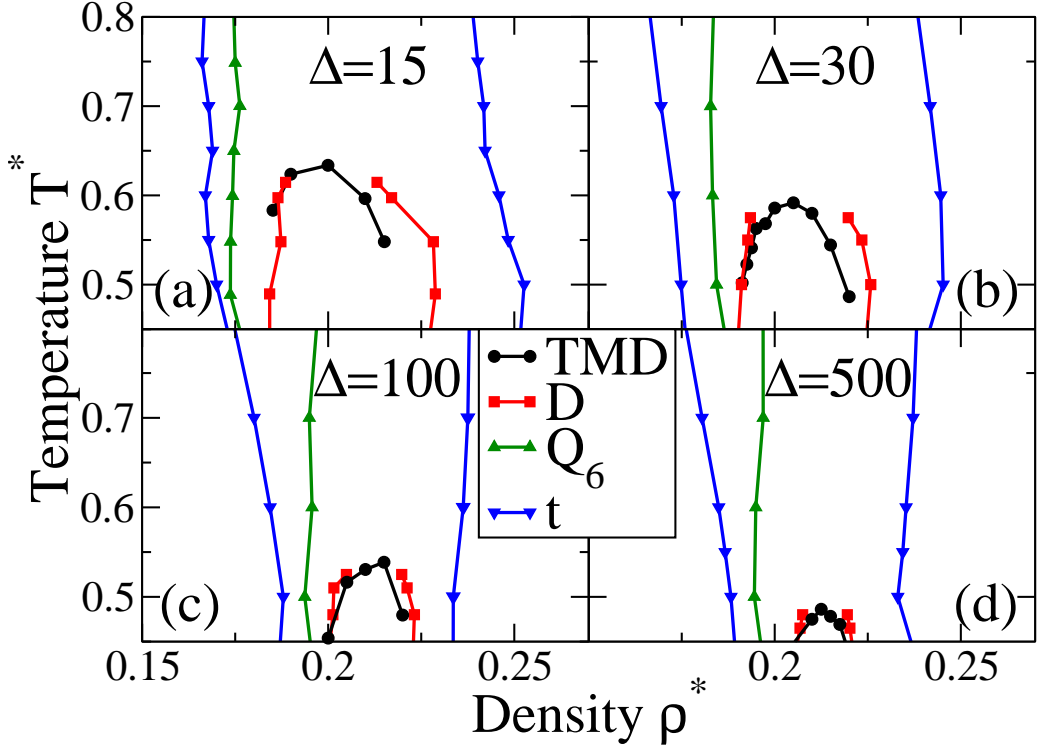}
\end{center}
	\caption{Hierarchy of anomalies in the $T^{*}-\rho^{*}$ plane,
          plotted for (a)  $\Delta=15$, (b) $\Delta=30$, (c)
          $\Delta=100$ and (d) $\Delta=500$. As the value of $\Delta$
          increases, both the regions of density anomaly 
and
          diffusion anomaly 
contract, with the diffusion
          anomaly region always encompasses the TMD line. The region
          of structural anomaly 
is only weakly affected.} 
\label{f7}       
\end{figure}

By varying the values of $\Delta$, the regions of anomalies are
affected in different ways while the hierarchy in which they occur is
always preserved (Fig.~\ref{f7}). 
We find that the anomalous region where $\rho$ decreases for
decreasing $T$ shrinks for increasing $\Delta$ and 
possibly tends to collapse onto one single point in the $P^{*}-T^{*}$
plane for $\Delta\rightarrow\infty$. This result would be consistent
with the behaviour of the DSW potential, that does not show density
anomaly \cite{Fra01,Fra02,Sk04,Ma05}.
Also the region of the diffusion anomaly
contracts as $\Delta$ increases. 
We find that the
diffusion anomaly region always encompasses the TMD line between
$\rho_{Dmin}$ and $\rho_{Dmax}$ (Fig.~\ref{f7}). 

The region of structural anomaly does not contract by increasing the value of $\Delta$ 
but tends asymptotically to a fixed region
in the $T^{*}-\rho^{*}$ plane (Fig.~\ref{f7}). 
This weak dependence on $\Delta$ suggests that the occurrence of
the structural anomaly does not disappear for very steep soft-core
potentials. 
This prediction  is consistent with the excess entropy
calculations 
that allow
de Oliveira et al. \cite{OFNB08} to argue
that the
structural anomaly  should be observable also for 
the DSW potential, here considered as the limit of the CSW for $\Delta
\rightarrow\infty$. 

\end{itemize}

\section{Outlook}

Isotropic soft-core potentials with two characteristic distances are
able to display the complex behaviour of anomalous liquids. They can
describe the effective interactions of systems such as colloids, protein solutions or
liquid metals. Due to the lack of directional interactions, these
potentials provide a mechanism for the anomalies that is alternative
to the bonding of network-forming liquid, e. g. water.
As a consequence, their use as coarse-grained models of water is
seriously disputed. In particular, they have a $P$-dependence of the structural
fluctuations and a supercooled phase diagram that are different from
those found in simulations of molecular models of water,
such as SPC/E, ST2, TIP4P, TIP5P (see for example \cite{brovchenko05})
or of other coarse-grained water models with 
explicit hydrogen bond interactions
\cite{KFS08,Mazza09,dlSantos09,Stokely2010}.
This wrong $P$-dependence affects, not only the supercooled state, but
also properties at ambient $T$, such as the velocity of sound. Water
sound propagation shows a discontinuity in $P$ that has a negative
slope in the $P$--$T$ phase diagram at 293~K and  0.29~GPa \cite{sound05},
consistent with the slope of the maxima of structural fluctuations found
at the same $T$ and $P$ 
in  numerical simulations of TIP4P-water by Saitta and Datchi
\cite{Saitta2003}.
Nevertheless, the isotropic soft-core potentials with two
characteristic distances display a sequence of anomalies that is water-like.

To elucidate their properties we have presented here the case of the
CSW potential whose anomalous behaviour is affected by the parameter
$\Delta$ associated to the steepness of the repulsive shoulder. 
For all the considered values of $\Delta$, the phase diagram displays
two first-order phase transitions, corresponding to a liquid-gas phase
transition at low densities and a liquid-liquid phase transition at
higher densities, both ending in critical points.
For  $15\leq \Delta \leq 500$ we verify that 
the anomalies in density, diffusion and structure are in the same
hierarchy as in water. We show that, as the value of
$\Delta$ increases, the regions of density and diffusion anomaly
contract in the $T-\rho$ plane, while the region of structural anomaly
is weakly affected. 

Since by increasing $\Delta$ the CSW potential approaches the
discontinuous shouldered well potential (DSW), the contraction of the
density anomaly and diffusion anomaly for $\Delta\rightarrow\infty$
is consistent with the fact that for the DSW potential no TMD is
observed. Our results suggest that the structural anomalies, instead,
should be present also in the DSW potential.
Therefore, the isotropic soft-core
potential with two characteristic distances can be considered as the
prototype for anomalous liquids with no directional bonding.
The fact that its
properties depend in a dramatic way on the details of the potential
calls for further investigations to understand better the mechanisms
that regulates the appearance of the anomalies.

\section*{Acknowledgments}

We thank MICINN-FEDER (Spain) grant FIS2009-10210 for support.

\bibliographystyle{aip}
\bibliography{vilaseca-franzese.bbl}

\begin{thebibliography}{100}

\bibitem{LJ}
J.~E. Lennard-Jones,
\newblock Proc. Camb. Phil. Soc. {\bf 27}, 469 (1931).

\bibitem{SY76}
M.~Silbert and W.~H. Young,
\newblock Phys. Lett. A {\bf 58}, 469 (1976).

\bibitem{YoungAlder77}
D.~A. Young and B.~J. Alder,
\newblock Phys. Rev. Lett. {\bf 38}, 1213 (1977).

\bibitem{KS77}
J.~M. Kincaid and G.~Stell,
\newblock J. Chem. Phys. {\bf 67}, 420 (1977).

\bibitem{LW77}
D.~Levesque and J.~J. Weis,
\newblock Phys. Lett. A {\bf 60}, 473 (1977).

\bibitem{KS78}
J.~M. Kincaid and G.~Stell,
\newblock Phys. Lett. A {\bf 65}, 131 (1978).

\bibitem{MAC79}
K.~K. Mon, N.~W. Ashcroft, and G.~V. Chester,
\newblock Phys. Rev. B {\bf 19}, 5103 (1979).

\bibitem{Cu81}
P.~T. Cummings and G.~Stell,
\newblock Mol. Phys. {\bf 43}, 1267 (1981).

\bibitem{VPRL83}
A.~Voronel, I.~Paperno, S.~Rabinovich, and E.~Lapina,
\newblock Phys. Rev. Lett. {\bf 50}, 247 (1983).

\bibitem{KH84}
G.~Kahl and J.~Hafner,
\newblock Solid State Comm. {\bf 49}, 1125 (1984).

\bibitem{DRB91}
P.~G. Debenedetti, V.~S. Raghavan, and S.~Borick,
\newblock J. Phys. Chem. {\bf 95}, 4540 (1991).

\bibitem{DebMetaLiq}
P.~G. Debenedetti,
\newblock {\em Metastable liquids: concepts and principles},
\newblock Princeton University Press, 7th edition ed., 1998.

\bibitem{SH93}
F.~H. Stillinger and T.~Head-Gordon,
\newblock Phys. Rev. E {\bf 47}, 2484 (1993).

\bibitem{Ve00}
E.~Velasco, L.~Medeiros, G.~Navascu\'es, P.~C. Hemmer, and G.~Stell,
\newblock Phys. Rev. Lett. {\bf 85}, 122 (2000).

\bibitem{BCESB00}
S.~H. Behrens, D.~I. Christl, R.~Emmerzael, P.~Schurtenberger, and M.~Borkovec,
\newblock Langmuir {\bf 16}, 2566 (2000).

\bibitem{Likos2001}
C.~N. Likos,
\newblock Physics Reports {\bf 348}, 267 (2001).

\bibitem{Quesada2001}
M.~Quesada-Perez, A.~Moncho-Jorda, F.~Martinez-Lopez, and R.~Hidalgo-Alvarez,
\newblock J. Chem. Phys. {\bf 115}, 10897 (2001).

\bibitem{Yethiraj03}
A.~Yethiraj and A.~van Blaaderen,
\newblock Nature (London) {\bf 421}, 513 (2003).

\bibitem{BAKSH04}
M.~M. Baksh, M.~Jaros, and J.~T. Groves,
\newblock Nature (London) {\bf 427}, 139 (2004).

\bibitem{protein}
S.~Brandon, P.~Katsonis, and P.~G. Vekilov,
\newblock Phys. Rev. E {\bf 73}, 061917 (2006).

\bibitem{Vekilov2008}
P.~G. Vekilov, W.~Pan, O.~Gliko, P.~Katsonis, and O.~Galkin,
\newblock {\em Aspects of Physical Biology}, volume 752 of {\em Lecture Notes
  in Physics}, chapter Metastable Mesoscopic Phases in Concentrated Protein
  Solutions, pages 65--95,
\newblock Springer Berlin / Heidelberg, 2008.

\bibitem{Buldyrev2008}
S.~Buldyrev,
\newblock {\em Aspects of Physical Biology}, volume 752 of {\em Lecture Notes
  in Physics}, chapter Application of Discrete Molecular Dynamics to Protein
  Folding and Aggregation, pages 97--131,
\newblock Springer Berlin / Heidelberg, 2008.

\bibitem{Bu02}
S.~V. Buldyrev, G.~Franzese, N.~Giovambattista, G.~Malescio, M.~R.
  Sadr-Lahijany, A.~Scala, A.~Skibinsky, and H.~E. Stanley,
\newblock Physica A {\bf 304}, 23 (2002).

\bibitem{Malescio07}
G.~Malescio,
\newblock J. Phys.: Condens. Matter {\bf 19} (2007).

\bibitem{KumarFBS2008}
P.~Kumar, G.~Franzese, S.~V. Buldyrev, and H.~E. Stanley,
\newblock {\em Aspects of Physical Biology}, volume 752 of {\em Lecture Notes
  in Physics}, chapter Dynamics of Water at Low Temperatures and Implications
  for Biomolecules, pages 3--22,
\newblock Springer Berlin / Heidelberg, 2008.

\bibitem{LSK76}
P.~Lamparter, S.~Stieb, and W.~Knoll,
\newblock Z. Naturforsch A {\bf 31}, 90 (1976).

\bibitem{Th76}
H.~Thurn and J.~Ruska,
\newblock J. Non-Cryst. Solids {\bf 22}, 331 (1976).

\bibitem{Sa67}
G.~E. Sauer and L.~B. Borst,
\newblock Science {\bf 158}, 1567 (1967).

\bibitem{Ke83}
S.~J. Kennedy and J.~C. Wheeler,
\newblock J. Chem. Phys. {\bf 78}, 1523 (1983).

\bibitem{Wa00}
J.~F. Wax, R.~Albaki, and J.~L. Bretonnet,
\newblock Phys. Rev. B {\bf 62}, 14818 (2000).

\bibitem{Ts91}
T.~Tsuchiya,
\newblock J. Phys. Soc. Jpn. {\bf 60}, 227 (1991).

\bibitem{An00}
C.~A. Angell, R.~D. Bressel, M.~Hemmatti, E.~J. Sare, and J.~C. Tucker,
\newblock Phys. Chem. Chem. Phys. {\bf 2}, 1559 (2000).

\bibitem{Ru06b}
R.~Sharma, S.~N. Chakraborty, and C.~Chakravarty,
\newblock J. Chem. Phys. {\bf 125}, 204501 (2006).

\bibitem{Sh02}
M.~S. Shell, P.~G. Debenedetti, and A.~Z. Panagiotopoulos,
\newblock Phys. Rev. E {\bf 66}, 011202 (2002).

\bibitem{Po97}
P.~H. Poole, M.~Hemmati, and C.~A. Angell,
\newblock Phys. Rev. Lett. {\bf 79}, 2281 (1997).

\bibitem{Sa03}
S.~Sastry and C.~A. Angell,
\newblock Nature Mater. {\bf 2}, 739 (2003).

\bibitem{FS07}
G.~Franzese and H.~E. Stanley,
\newblock J. Phys.: Condens. Matter {\bf 19}, 205126 (2007).

\bibitem{An76}
C.~A. Angell, E.~D. Finch, and P.~Bach,
\newblock J. Chem. Phys. {\bf 65}, 3065 (1976).

\bibitem{St83}
P.~J. Steinhardt, D.~R. Nelson, and M.~Ronchetti,
\newblock Phys. Rev. B {\bf 28}, 784 (1983).

\bibitem{Er01}
J.~R. Errington and P.~D. Debenedetti,
\newblock Nature (London) {\bf 409}, 318 (2001).

\bibitem{Er03}
J.~E. Errington, P.~G. Debenedetti, and S.~Torquato,
\newblock J. Chem. Phys. {\bf 118}, 2256 (2003).

\bibitem{Po92}
P.~H. Poole, F.~Sciortino, U.~Essmann, and H.~E. Stanley,
\newblock Nature (London) {\bf 360}, 324 (1992).

\bibitem{Mishima1985}
O.~Mishima, L.~Calvert, and E.~Whalley,
\newblock Nature (London) {\bf {314}}, {76 ({1985}).

\bibitem{VHDA}
T.~Loerting, C.~Salzmann, I.~Kohl, E.~Mayer, and A.~Hallbrucker,
\newblock Phys. Chem. Chem. Phys. {\bf {3}}, {5355 ({2001}).

\bibitem{polymorphysmRev}
M.~C. Wilding, M.~Wilson, and P.~F. McMillan,
\newblock Chem. Soc. Rev. {\bf 35}, 964  (2006).

\bibitem{Ka00}
Y.~Katayama, T.~Mizutani, W.~Utsumi, O.~Shimomura, M.~Yamakata, and
  K.~Funakoshi,
\newblock Nature (London) {\bf 403}, 170 (2000).

\bibitem{Ka04}
Y.~Katayama, Y.~Inamura, T.~Mizutani, W.~Yamakata, M.~Utsumi, and S.~O.,
\newblock Science {\bf 306}, 848 (2004).

\bibitem{Mo03}
G.~Monaco, F.~S., W.~A. Crichton, and M.~Mezouar,
\newblock Phys. Rev. Lett. {\bf 90}, 255701 (2003).

\bibitem{ReiKurita10292004}
R.~Kurita and H.~Tanaka,
\newblock Science {\bf 306}, 845 (2004).

\bibitem{Ta04}
H.~Tanaka, R.~Kurita, and H.~Mataki,
\newblock Phys. Rev. Lett. {\bf 92}, 025701 (2004).

\bibitem{Kur05}
R.~Kurita and H.~Tanaka,
\newblock J. Phys.: Condens. Matter {\bf 17}, L293 (2005).

\bibitem{Greaves08}
G.~N. Greaves, M.~C. Wilding, S.~Fearn, D.~Langstaff, F.~Kargl, S.~Cox, Q.~V.
  Van, O.~Majerus, C.~J. Benmore, R.~Weber, C.~M. Martin, and L.~Hennet,
\newblock Science {\bf {322}}, {566 ({2008}).

\bibitem{An96}
C.~A. Angell, S.~Borick, and M.~Grabow,
\newblock J. Non-Cryst. Solids {\bf 207}, 463 (1996).

\bibitem{La00}
D.~J. Lacks,
\newblock Phys. Rev. Lett. {\bf 84}, 4629 (2000).

\bibitem{Th93}
M.~van Thiel and F.~H. Ree,
\newblock Phys. Rev. B {\bf 48}, 3591 (1993).

\bibitem{Br98}
V.~V. Brazhkin, E.~L. Gromnistkaya, O.~V. Stalgorova, and A.~G. Lyapin,
\newblock Rev. High Pressure Sci. Technol. {\bf 7}, 1129 (1998).

\bibitem{Va03}
M.~G. Vasin and V.~I. Lad\'yanov,
\newblock Phys. Rev. E {\bf 68}, 051202 (2003).

\bibitem{Aa94}
S.~Aasland and P.~F. McMillan,
\newblock Nature (London) {\bf 369}, 633 (1994).

\bibitem{Wi01}
M.~C. Wilding and P.~F. McMillan,
\newblock J. Non-Cryst. Solids {\bf 293}, 357 (2001).

\bibitem{WBM02}
M.~C. Wilding, C.~J. Benmore, and P.~F. McMillan,
\newblock J. Non-Cryst. Solids {\bf 297}, 143 (2002).

\bibitem{brovchenko05}
I.~Brovchenko, A.~Geiger, and A.~Oleinikova,
\newblock J. Chem. Phys. {\bf 123}, 044515 (2005).

\bibitem{Mo01}
T.~Morishita,
\newblock Phys. Rev. Lett. {\bf 87}, 4659 (2001).

\bibitem{Vo01b}
F.~Saika-Voivod, F.~Sciortino, and P.~H. Poole,
\newblock Phys. Rev. E {\bf 63}, 011202 (2001).

\bibitem{Sca03}
S.~Scandolo,
\newblock Proc. Natl. Acad. Sci. U.S.A. {\bf 100}, 3051 (2003).

\bibitem{Gl99}
J.~N. Glosli and F.~H. Ree,
\newblock Phys. Rev. Lett. {\bf 82}, 4659 (1999).

\bibitem{Christine02}
C.~J. Wu, J.~N. Glosli, G.~Galli, and F.~H. Ree,
\newblock Phys. Rev. Lett. {\bf 89}, 135701 (2002).

\bibitem{Gh05}
L.~M. Ghiringhelli, J.~H. Los, E.~J. Meijer, A.~Fasolino, and D.~Frenkel,
\newblock Phys. Rev. Lett. {\bf 94}, 145701 (2005).

\bibitem{HemmerStellA}
P.~C. Hemmer and G.~Stell,
\newblock Phys. Rev. Lett. {\bf 24}, 1284 (1970).

\bibitem{St72}
G.~Stell and P.~C. Hemmer,
\newblock J. Chem. Phys. {\bf 56}, 4274 (1972).

\bibitem{Ja98}
E.~A. Jagla,
\newblock Phys. Rev. E {\bf 58}, 1478 (1998).

\bibitem{Ja99a}
E.~A. Jagla,
\newblock J. Chem. Phys. {\bf 110}, 451 (1999).

\bibitem{Ja99b}
E.~A. Jagla,
\newblock J. Chem. Phys. {\bf 111}, 8980 (1999).

\bibitem{Ja01a}
E.~A. Jagla,
\newblock Phys. Rev. E {\bf 63}, 061501 (2001).

\bibitem{Ja01b}
E.~A. Jagla,
\newblock Phys. Rev. E {\bf 63}, 061509 (2001).

\bibitem{Wi02}
N.~B. Wilding and J.~E. Magee,
\newblock Phys. Rev. E {\bf 66}, 031509 (2002).

\bibitem{Ku05}
P.~Kumar, S.~V. Buldyrev, F.~Sciortino, E.~Zaccarelli, and H.~E. Stanley,
\newblock Phys. Rev. E {\bf 72}, 021501 (2005).

\bibitem{Yan05}
Z.~Yan, S.~V. Buldyrev, N.~Giovambattista, and H.~E. Stanley,
\newblock Phys. Rev. Lett. {\bf 95}, 130604 (2005).

\bibitem{Yan06}
Z.~Yan, S.~V. Buldyrev, N.~Giovambattista, P.~G. Debenedetti, and H.~E.
  Stanley,
\newblock Phys. Rev. E {\bf 73}, 051204 (2006).

\bibitem{Yan08}
Z.~Yan, S.~V. Buldyrev, P.~Kumar, N.~Giovambattista, and H.~E. Stanley,
\newblock Phys. Rev. E {\bf 77}, 042201 (2008).

\bibitem{Wi06}
H.~M. Gibson and N.~B. Wilding,
\newblock Phys. Rev. E {\bf 73}, 061507 (2006).

\bibitem{Ol06a}
A.~B. de~Oliveira, P.~A. Netz, T.~Colla, and M.~C. Barbosa,
\newblock J. Chem. Phys. {\bf 124}, 084505 (2006).

\bibitem{Ol06b}
A.~B. de~Oliveira, P.~A. Netz, T.~Colla, and M.~C. Barbosa,
\newblock J. Chem. Phys. {\bf 125}, 124503 (2006).

\bibitem{Ki76a}
J.~M. Kincaid, G.~Stell, and C.~K. Hall,
\newblock J. Chem. Phys. {\bf 65}, 2161 (1976).

\bibitem{Ki76b}
J.~M. Kincaid, G.~Stell, and E.~Goldmark,
\newblock J. Chem. Phys. {\bf 65}, 2172 (1976).

\bibitem{Ha73}
C.~K. Hall and G.~Stell,
\newblock Phys. Rev. A {\bf 7}, 1679 (1973).

\bibitem{La98}
M.~R. Sadr-Lahijany, A.~Scala, S.~V. Buldyrev, and H.~E. Stanley,
\newblock Phys. Rev. Lett. {\bf 81}, 4895 (1998).

\bibitem{La99}
M.~R. Sadr-Lahijany, A.~Scala, S.~V. Buldyrev, and H.~E. Stanley,
\newblock Phys. Rev. E {\bf 60}, 6714 (1999).

\bibitem{Sc00}
A.~Scala, M.~R. Sadr-Lahijany, N.~Giovambattista, S.~V. Buldyrev, and H.~E.
  Stanley,
\newblock J. Stat. Phys. {\bf 100}, 97 (2000).

\bibitem{Sc01}
A.~Scala, M.~R. Sadr-Lahijany, N.~Giovambattista, S.~V. Buldyrev, and H.~E.
  Stanley,
\newblock Phys. Rev. E {\bf 63}, 041202 (2001).

\bibitem{Fra01}
G.~Franzese, G.~Malescio, A.~Skibinsky, S.~V. Buldyrev, and H.~E. Stanley,
\newblock Nature (London) {\bf 409}, 692 (2001).

\bibitem{Ma05}
G.~Malescio, G.~Franzese, A.~Skibinsky, S.~V. Buldyrev, and H.~E. Stanley,
\newblock Phys. Rev. E {\bf 71}, 061504 (2005).

\bibitem{Bu03}
S.~V. Buldyrev and H.~E. Stanley,
\newblock Physica A {\bf 330}, 124 (2003).

\bibitem{BB04}
A.~L. Balladares and M.~C. Barbosa,
\newblock Journal of Physics: Condensed Matter {\bf 16}, 8811 (2004).

\bibitem{Ol05}
A.~B. de~Oliveira and M.~C. Barbosa,
\newblock J. Phys.: Condens. Matter {\bf 17}, 399 (2005).

\bibitem{Camp03}
P.~J. Camp,
\newblock Phys. Rev. E {\bf 68}, 061506 (2003).

\bibitem{Camp05}
P.~J. Camp,
\newblock Phys. Rev. E {\bf 71}, 031507 (2005).

\bibitem{Fr07a}
G.~Franzese,
\newblock J. Mol. Liq. {\bf 136}, 267 (2007).

\bibitem{OFNB08}
A.~B. de~Oliveira, G.~Franzese, P.~A. Netz, and M.~C. Barbosa,
\newblock J. Chem. Phys. {\bf 128}, 064901 (2008).

\bibitem{GFFR09}
N.~V. Gribova, Y.~D. Fomin, D.~Frenkel, and V.~N. Ryzhov,
\newblock Phys. Rev. E {\bf 79}, 051202 (2009).

\bibitem{BeeSw}
R.~I. Beecroft and C.~A. Swenson,
\newblock J. Phys. Chem. Solids {\bf 15}, 234 (1960).

\bibitem{HG93}
T.~Head-Gordon and F.~H. Stillinger,
\newblock J. Chem. Phys. {\bf 98}, 3313 (1993).

\bibitem{SS97}
F.~H. Stillinger and D.~K. Stillinger,
\newblock Physica A {\bf 244}, 385 (1997).

\bibitem{HydraProBi}
M.~C. Bellissent-Funel, editor,
\newblock {\em Hydration Processes in Biology. Theoretical and Experimental
  Aproaches}, volume 305 of {\em NATO Advanced Studies Institute, Series A:
  Life Sciences},
\newblock IOS Press, 1998.

\bibitem{RobSinZhuEv}
G.~W. Robinson, S.~Singh, S.-B. Zhu, and M.~W. Evans,
\newblock {\em Water in Biology, Chemistry and Physics},
\newblock World Scientific, 1996.

\bibitem{SW78}
F.~H. Stillinger and T.~A. Weber,
\newblock J. Chem. Phys. {\bf 68}, 3837 (1978).

\bibitem{Archer07}
A.~J. Archer and N.~B. Wilding,
\newblock Phys. Rev. E {\bf 76}, 031501 (2007).

\bibitem{Ch96b}
C.~H. Cho, S.~Singh, and G.~W. Robinson,
\newblock Faraday Discuss. {\bf 103}, 19 (1996).

\bibitem{Ch96a}
C.~H. Cho, S.~Singh, and G.~W. Robinson,
\newblock Phys. Rev. Lett. {\bf 76}, 1651 (1996).

\bibitem{Ne04}
P.~A. Netz, J.~F. Raymundi, A.~S. Camera, and M.~C. Barbosa,
\newblock Physica A {\bf 342}, 48 (2004).

\bibitem{Qui05}
D.~Quigley and M.~I.~J. Probert,
\newblock Phys. Rev. E {\bf 71}, 065701 (2005).

\bibitem{He01}
P.~C. Hemmer, E.~Velasco, L.~Medeiros, G.~Navascu\'es, and G.~Stell,
\newblock J. Chem. Phys. {\bf 114}, 2268 (2001).

\bibitem{Xu05}
L.~Xu, P.~Kumar, S.~V. Buldyrev, S.-H. Chen, P.~Poole, F.~Sciortino, and H.~E.
  Stanley,
\newblock Proc. Natl. Acad. Sci. U.S.A. {\bf 102}, 16558 (2005).

\bibitem{Xu06}
L.~Xu, S.~Buldyrev, C.~A. Angell, and H.~E. Stanley,
\newblock Phys. Rev. E {\bf 74}, 031108 (2006).

\bibitem{Ca06}
J.~B. Caballero and A.~M. Puertas,
\newblock Phys. Rev. E {\bf 74}, 051506 (2006).

\bibitem{Fra02}
G.~Franzese, G.~Malescio, A.~Skibinsky, S.~V. Buldyrev, and H.~E. Stanley,
\newblock Phys. Rev. E {\bf 66}, 051206 (2002).

\bibitem{Sk04}
A.~Skibinsky, S.~V. Buldyrev, G.~Franzese, G.~Malescio, and H.~E. Stanley,
\newblock Phys. Rev. E {\bf 69}, 061206 (2004).

\bibitem{rzysko08}
W.~Rzysko, O.~Pizio, A.~Patrykiejew, and S.~Sokolowski,
\newblock J. Chem. Phys. {\bf 129}, 124502 (2008).

\bibitem{cervantes07}
L.~A. Cervantes, A.~L. Benavides, and F.~del R\'{\i}o,
\newblock J. Chem. Phys. {\bf 126}, 084507 (2007).

\bibitem{Artemenko08}
S.~Artemenko, T.~Lozovsky, and V.~Mazur,
\newblock J. Phys.: Condens. Matter {\bf 20}, 44119 (2008).

\bibitem{Loerting_Giovambattista}
T.~Loerting and N.~Giovambattista,
\newblock J. Phys: Cond. Mat. {\bf 18}, R919 (2006).

\bibitem{MalPel2003}
G.~Malescio and G.~Pellicane,
\newblock Nature Mater. {\bf 2}, 97 (2003).

\bibitem{duncan04}
P.~D. Duncan and P.~J. Camp,
\newblock J. Chem. Phys. {\bf 121}, 11322 (2004).

\bibitem{Osterman07}
N.~Osterman, D.~Babi\ifmmode~\check{c}\else \v{c}\fi{}, I.~Poberaj,
  J.~Dobnikar, and P.~Ziherl,
\newblock Phys. Rev. Lett. {\bf 99}, 248301 (2007).

\bibitem{Dobnikar08}
J.~Dobnikar, J.~Fornleitner, and G.~Kahl,
\newblock J. Phys.: Condens. Matte {\bf 20}, 494220 (2008).

\bibitem{Pizio2009a}
O.~Pizio, H.~Dominguez, Y.~Duda, and S.~Soko{\l}owski,
\newblock J. Chem. Phys. {\bf 130}, 174504 (2009).

\bibitem{Hafner90}
J.~Hafner,
\newblock J. Phys.: Condens. Matter {\bf 2}, 1271 (1990).

\bibitem{Standaert10}
S.~Standaert, J.~Ryckebusch, and L.~De~Cruz,
\newblock Creating conditions of anomalous self-diffusion in a liquid with
  molecular dynamics (preprint arxiv:0907.1856v2),
\newblock 2009.

\bibitem{Vilaseca10}
P.~Vilaseca and G.~Franzese,
\newblock Softness dependence of the anomalies for the continuous shouldered
  well potential (preprint arxiv:1004.3186v1),
\newblock 2010.

\bibitem{KFS08}
P.~Kumar, G.~Franzese, and H.~E. Stanley,
\newblock Phys. Rev. Lett. {\bf 100}, 105701 (2008).

\bibitem{Mazza09}
M.~G. Mazza, K.~Stokely, E.~G. Strekalova, H.~E. Stanley, and G.~Franzese,
\newblock Comp. Phys. Comm. {\bf 180}, 497 (2009).

\bibitem{dlSantos09}
F.~d.~l. Santos and G.~Franzese,
\newblock Influence of intramolecular couplings in a model for hydrogen-bonded
  liquids,
\newblock volume 1091, pages 185--197, Granada (Spain), 2009, AIP.

\bibitem{Stokely2010}
K.~Stokely, M.~G. Mazza, H.~E. Stanley, and G.~Franzese,
\newblock Proc. Natl. Acad. Sci. U.S.A. {\bf 107}, 1301 (2010).

\bibitem{sound05}
F.~Li, Q.~Cui, Z.~He, T.~Cui, J.~Zhang, Q.~Zhou, G.~Zou, and S.~Sasaki,
\newblock J. Chem. Phys. {\bf 123}, 174511 (2005).

\bibitem{Saitta2003}
A.~M. Saitta and F.~Datchi,
\newblock Phys. Rev. E {\bf 67}, 020201 (2003).
}}}
\end{thebibliography}

\end{document}